\documentclass[fleqn,10pt]{wlscirep}
\title{Quantum Zeno and Zeno-like effects in nitrogen vacancy centers}

\author[1,+]{Jing Qiu}
\author[1,+]{Yang-Yang Wang}
\author[2]{Zhang-Qi Yin}
\author[1]{Mei Zhang}
\author[1,*]{Qing Ai}
\author[1]{Fu-Guo Deng}

\affil[1]{Department of Physics, Applied Optics Beijing Area Major Laboratory,
Beijing Normal University, Beijing 100875, China}

\affil[2]{Center for Quantum Information, Institute for Interdisciplinary Information Sciences, Tsinghua University, Beijing 100084, China}

\affil[*]{aiqing@bnu.edu.cn}

\affil[+]{these authors contributed equally to this work}


\begin{abstract}
We present a proposal to realize the quantum Zeno effect (QZE) and quantum Zeno-like effect (QZLE) in a proximal $\mathrm{^{13}C}$ nuclear spin by controlling a proximal electron spin of a nitrogen vacancy (NV) center. The measurement is performed by applying a microwave pulse to induce the transition between different electronic spin states. Under the practical experimental conditions, our calculations show that there exist both QZE and QZLE in a $^{13}$C nuclear spin in the vicinity of an NV center.
\end{abstract}

\begin{document}

\flushbottom
\maketitle
\thispagestyle{empty}

%\section*{Introduction}

Quantum Zeno effect \cite{Khalfin68,Misra77} (QZE) is a very interesting phenomenon in quantum  physics, in which the evolution of a quantum system can be inhibited by
frequent measurements. In 1990, based on Cook's theoretical
proposal \cite{Cook88},  QZE \cite{Itano90} was experimentally demonstrated
by controlling the transition between two hyperfine levels of
$\mathrm{^{9}Be^{+}}$ with laser pulses, and there was a good agreement
between theoretical prediction and experimental results.
Since then, much effort has been paid to the research on
QZE \cite{Wang08,Xu11,Bernu08,Zhang15}, quantum anti-Zeno effect \cite{Kofman00,Facchi01,Ai10,Ai13}
and quantum Zeno-like effect (QZLE) \cite{Layden15,Nakazato03}.
Because of its potential application in slowing down or even freezing
the dynamic evolution of a system via repeated
frequent measurements, it recently has attracted considerable interest
as a tool in the fields of quantum information processing \cite{Xue15,Hua15,Liu11,Li11,Li12,Yu12,Liu13}
and ultrasensitive magnetometer \cite{Wang13}.

QZE has been successfully demonstrated on
various physical systems, such as trapped ions \cite{Balzer00},
superconducting qubits \cite{Matsuzaki10,Zhang11}, cavity quantum electrodynamics
\cite{Bernu08,Hua15}, nuclear magnetic resonance \cite{Xiao06,Alvarez10},
and Bose-Einstein condensates \cite{Streed06,Bar-Gill09,Zhang03}. On the other hand,
in solid-state quantum-information technology, a
nitrogen vacancy (NV) center which consists of a nitrogen
substituting for a carbon and an adjacent vacancy in
diamond has been identified as one of the most promising candidates for qubits \cite{Zhou14,Yang10,WeiNVgate,RenNVhyperCoup,ZhangMPRA,Song15,Kost15,Yin15} due
to its long coherence time at room temperature \cite{Kennedy03,Bar-Gill13,Maurer12}
and convenient manipulation under optical field, microwave field and rf field
\cite{Dutt07,Zu14}. Quantum Zeno-like phenomenon was experimentally
demonstrated by inhibiting coherent spin dynamics induced by the microwave driving
between two ground-state electron-spin levels ($m_s=0$ and $m_s=1$) of a single NV center \cite{Wolters13}.
Therein, only one measurement is performed to analyze the measurement effect on
electron-spin states, i.e., the population variation between the electron-spin states $m_s=0$ and $m_s=1$ has been made by this single measurement.

%The limit of infinite measurements is almost impossible for the experimental situation.

In the conventional QZE, repeated instantaneous perfect measurements
performed on the system will freeze the evolution of the initial state.
The perfect conventional QZE requires infinite measurements with repetitive frequency
approaching infinity, which might be impossible in experiment.
However, it was recently discovered that perfect freezing of quantum
states can also be achieved by more realistic non-projective measurements
performed at a finite frequency \cite{Layden15}. According to Brouwer's
fixed-point theorem \cite{Brouwer11}, there always exist some quantum states
which satisfy $\Phi(\rho_0)=\rho_0$, where $\Phi$ represents a quantum dynamical
evolution process of a system with its initial state $\rho_0$.
After \emph{n} identical cycle process, the system stays at the state as the same as its inital one, i.e., $\Phi^n(\rho_0)=\rho_0$. In this way, a QZLE can be achieved
with finite-frequency measurements.

In this paper, inspired by the discovery in Ref.~\citen{Layden15},
we present a proposal to achieve the QZLE in a proximal $\mathrm{^{13}C}$ nuclear spin of
an NV center by controlling the electron spin.
In our proposal, the electron spin plays the role as a detector
while the $\mathrm{^{13}C}$ nuclear spin acts as the target. Furthermore,
the conventional QZE is demonstrated by modulating the measurement
parameters and the external magnetic field. Here, instead of projective measurements,
we apply a microwave pulse to induce the transition between
different electronic states, followed by initialization of electron spin.
Our numerical calculation properly shows that for suitable parameters there exist the
conventional QZE and QZLE in a proximal $\mathrm{^{13}C}$ nuclear spin of the NV center.

\section*{Results}

\subsection*{The model}

An NV center in diamond is composed of a nitrogen atom and a vacancy in an adjacent lattice site.
It is a defect with $\mathrm{C}_{3\mathrm{v}}$ symmetry \cite{Maze11,Doherty11}.
For the negatively-charged NV center with electron spin \emph{S}=1, the ground state is a spin-triplet state $\mathrm{^{3}A}$ with a zero-field splitting \emph{D}=2.87 GHz between spin sublevels $m_{s}=0$ and $m_{s}=\pm1$ \cite{Jacques09}.
Around NV centers there are three kinds of nuclear spins \cite{Smeltzer09,Felton09}, i.e.,
$^\textrm{13}\textrm{C}$ (\emph{I}=1/2), $^\textrm{14}\textrm{N}$ (\emph{I}=1),
and $^\textrm{15}\textrm{N}$ (\emph{I}=1/2). They can be manipulated by microwave and rf fields.

%In this article, we realize the QZE by controlling

%In this proposal, the electron spin serves as the detector and the nuclear spin
%acts as the target.

\begin{figure}[ht]
\centering
\includegraphics[bb=0bp 20bp 650bp 370bp,scale=0.4,angle=0]{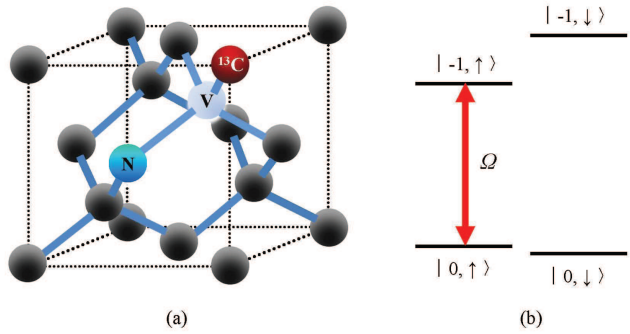}
\caption{Scheme for demonstration of the QZE in an NV center. (a) A $^{13}\mathrm{C}$ nuclear spin is
at the nearest-neighbor lattice site of an NV center.
(b) The energy-level diagram of the ground state
hyperfine structure where a microwave drives the transition between
$\left|0,\uparrow\right\rangle$ and $\left|-1,\uparrow\right\rangle$
with Rabi frequency $\varOmega$ and driving frequency $\omega$.\label{fig:Scheme}}
\label{fig:stream}
\end{figure}

Consider an NV center and a $^{\mathrm{13}}\mathrm{C}$ nuclear spin
which locates in the first coordination shell around the
NV center \cite{Jelezko04}, as shown in Fig.~\ref{fig:Scheme}(a).
In other words, this $^{\mathrm{13}}\mathrm{C}$ nuclear spin is at the
nearest-neighbor lattice site of the NV center.
As a result, there is a strong hyperfine coupling between the nuclear and electronic spins.
Figure~\ref{fig:Scheme}(b) shows the simplified energy-level
diagram of the ground-state hyperfine structure associated with the nearest-neighbor
$^{13}\mathrm{C}$ nuclear spin. To demonstrate the QZE, the electron-spin
states ($m_{s}=\!-1,~0$) and nuclear-spin states ($|\!\uparrow\rangle,~|\!\downarrow\rangle$)
are chosen to code qubits. The target and detector are initially uncorrelated,
i.e., they are in a product state. A strong electron-spin polarization into
the $m_{s}=0$ sublevel can be induced by circulatory optical excitation-emission.
This effect results from spin-selective non-radiative intersystem crossing to
a metastable state lying between the ground and excited triplet states \cite{Tamarat08,Manson06}.
Moreover, the nuclear spin could be well isolated from the electron spin, during the optical polarization and measurement of the electronic state \cite{Dutt07,Jiang08}. In other words, the state of nuclear spin could be unperturbed when the initialization and measurement are performed on the electronic spin.

Suppose that the electron spin is initially in
its ground state $|0\rangle$ and the nuclear spin is in an arbitrary state.
First of all, the whole system evolves freely for a time interval $\varDelta t_{f}$.
Afterwards, a microwave driving is used to perform measurement.
As Fig.~\ref{fig:Scheme}(b) shows, the microwave drives
the transition between $\left|0,\uparrow\right\rangle$ and $\left|-1,\uparrow\right\rangle$
with Rabi frequency $\varOmega$ and driving frequency $\omega$.
In the process of measurement, the total system evolves under the Hamiltonian $H_{M}=H_{F}+H_{I}$
for a time interval $\varDelta t_{m}$, where $H_{F}$ is the free
Hamiltonian without measurement and $H_{I}$ is the interaction Hamiltonian
describing the transition induced by microwave driving. After the measurement, by optical pumping, the electron spin is initialized in its ground state $|0\rangle$, and meanwhile the electron and nuclear spins are decoupled \cite{Dutt07,Jiang08,van-der-Sar12}.
And then the above process is repeated. When the duration of the 532-nm light pulse for optical pumping is appropriate, the $^{13}\mathrm{C}$ nuclear spin could be well isolated from the electron
spin and the nuclear spin state can be preserved. In particular, the dephasing of nuclear spin
 can hardly be observed for light pulses of $\sim$\!140 ns which is sufficiently long to
polarize the electron spin while leave the state of nuclear spin undisturbed. \cite{Dutt07}

\subsubsection*{The free Hamiltonian $H_{F}$}

The general Hamiltonian of an NV center and a $^{13}\mathrm{C}$
nuclear spin which locates in the first coordination shell around the NV center is \cite{Dreau13}
\begin{align}
H =&~DS_{z}^{2}+\gamma_{e}\overrightarrow{B}\cdot\overrightarrow{S}+\gamma_{n}\overrightarrow{B}\cdot\overrightarrow{I}+\overrightarrow{S}A\overrightarrow{I} \\
	  =&~DS_{z}^{2}+\gamma_{e}(B_{x}S_{x}+B_{y}S_{y}+B_{z}S_{z})+\gamma_{n}(B_{x}I_{x}+B_{y}I_{y}+B_{z}I_{z}) +(A_{xx}S_{x}I_{x}+A_{yy}S_{y}I_{y}+A_{zz}S_{z}I_{z}). \nonumber
\end{align}
Here, the first term stands for the zero-field splitting of the electronic ground state.
The second term is the Zeeman energy splitting of the electron with $\gamma_{e}$
being the electronic gyromagnetic ratio. The third term denotes
the nuclear Zeeman effect where $\gamma_{n}$ is the $^{13}\mathrm{C}$
nuclear spin gyromagnetic ratio. And the last term describes the hyperfine interaction
between the electron spin and the nuclear spin of $^{13}\mathrm{C}$ atom.

Using a permanent magnet, an external magnetic field $B_{z}$ is applied parallel to the NV
axis. Hence, $\gamma_{e}(B_{x}S_{x}+B_{y}S_{y})$ and $\gamma_{n}(B_{x}I_{x}+B_{y}I_{y})$
are removed. Under the condition of weak magnetic field strength, the difference between
 the zero-field splitting $D=2.87$ GHz and the electronic Zeeman splitting is much larger than the hyperfine interaction. In this situation, the electron-nuclear-spin flip-flop processes induced by the hyperfine interaction are sufficiently suppressed. Therefore, this allows for the secular approximation \cite{Dreau13,Childress06,Neumann08}, and the $S_{x}I_{x}$ and $S_{y}I_{y}$ terms can be neglected.
In other words, for the weak external magnetic field along the NV axis
only the longitudinal hyperfine interaction needs to be taken into account, and the ground-state manifold of the NV center coupled with a proximal $^{13}\mathrm{C}$
nuclear spin is described by the Hamiltonian
\begin{equation}
H_{F}=DS_{z}^{2}+\gamma_{e}B_{z}S_{z}+\gamma_{n}B_{z}I_{z}+A_{zz}S_{z}I_{z},
\end{equation}
where
\begin{align}
I_{z} & = \frac{1}{2}\sigma_{z},\\
S_{z} & = \left(
              \begin{array}{ccc}
                1 & 0 & 0 \\
                0 & 0 & 0 \\
                0 & 0 & -1 \\
              \end{array}
            \right),
\end{align}
and $\sigma_{z}$ is the Pauli-$z$ operator.

\subsubsection*{The Hamiltonian under measurement}

A microwave driving is utilized to perform the measurement,
as shown in Fig.~\ref{fig:Scheme}(b). This microwave
pulse drives the transition between $\left|0\right\rangle$ and $\left|-1\right\rangle$.
The driving frequency is set to be resonant with the transition
between $\left|0,\uparrow\right\rangle$ and $\left|-1,\uparrow\right\rangle$, and meanwhile largely
detuned from that between $\left|0,\downarrow\right\rangle$ and $\left|-1,\downarrow\right\rangle$.
In this way, the transition between $\left|0,\uparrow\right\rangle$
and $\left|-1,\uparrow\right\rangle$ can be induced selectively. Thus the system evolves under
the whole Hamiltonian $H_{M}=H_{F}+H_{I}$, where the interaction Hamiltonian is
\begin{equation}
H_{I}=\varOmega e^{i\omega t}|0,\uparrow\rangle\langle-1,\uparrow|+\textrm{h.c.}
\end{equation}

In order to analytically calculate the quantum dynamics under the influence of $H_M$, we transform to the rotating frame defined by the transformation
$\bigr|\varPsi(t)^{R}\bigl\rangle=U^{\dagger}(t)\bigr|\varPsi(t)\bigl\rangle$, where $U(t)=\exp(-i H_F t)$, $\bigr|\varPsi(t)\bigl\rangle$ and $\bigr|\varPsi(t)^{R}\bigl\rangle$ are the wave functions in the static and rotating frames respectively. Therefore, using equation~(\ref{eq:rotating-frame}),
the Hamiltonian under the rotating frame can be obtained, i.e.,
\begin{align}
H_{M}^{R}	&=	U^{\dagger}H_{M}U-H_F =	\varOmega |0,\uparrow\rangle\langle-1,\uparrow|+\textrm{h.c.}
\end{align}

\subsection*{Dynamic Evolution}

Now, let us demonstrate the quantum dynamics of the whole system in a full cycle,
which includes a free evolution followed by a measurement process.
The electron spin is initially prepared in its ground
state, i.e.,
\begin{equation}
\rho^{(D)}(0)=~\bigr|0\bigl\rangle\bigr\langle0\bigr|,
\end{equation}
and the nuclear spin is in an arbitrary state
\begin{equation}
\rho^{(T)}(0)=\left(\begin{array}{cc}
\alpha & \beta\\
\beta^{*} & 1-\alpha
\end{array}\right),
\end{equation}
which is spanned by the basis $\{\vert \uparrow\rangle ,\vert \downarrow\rangle \}$.
Thus, the initial state of the total system is $\rho(0)=\rho^{(D)}(0)\otimes\rho^{(T)}(0)$.

In the free evolution, the total system evolves under its free Hamiltonian
$H_{F}$ for a time interval $\varDelta t_{f}$, which is described by the
evolution operator
\begin{equation}
U_{f}(\varDelta t_{f})=e^{-iH_{F}\varDelta t_{f}}.
\end{equation}
Apparently, without the driving, the evolution operator of the total system
is diagonal. At the end of free evolution, the state
of the total system becomes $\rho(\varDelta t_{f})=U_{f}(\varDelta t_{f})\rho(0)U_{f}^{\dagger}(\varDelta t_{f})$.
Afterwards, a microwave pulse is used to drive the transition between
$\left|0,\uparrow\right\rangle$ and $\left|-1,\uparrow\right\rangle$.
The total system evolves under the Hamiltonian $H_{M}^{R}$ for a time interval
$\varDelta t_{m}$. Having transformed to the rotating frame, a time-independent Hamiltonian is obtained and the corresponding evolution operator is
\begin{equation}
U_{m}^{R}(\varDelta t_{m})=e^{-iH_{M}^{R}\varDelta t_{m}}. \label{eq:evolve-Hm}
\end{equation}
After a cycle with duration $\tau=\varDelta t_{f}+\varDelta t_{m}$, the
state of the whole system is
\begin{equation}
\rho^{R}(\tau)=U_{m}^{R}U_{f}\rho(0)U_{f}^{\dagger}U_{m}^{R\dagger}.
\end{equation}
Utilizing equation~(\ref{eq:rotating-frame-state}),
the final state of the whole system in the static frame reads
\begin{equation}
\rho(\tau)=U(\varDelta t_{m})U_{m}^{R}(\varDelta t_{m})U_{f}(\varDelta t_{f})\rho(0)U_{f}^{\dagger}(\varDelta t_{f})U_{m}^{R\dagger}(\varDelta t_{m})U^{\dagger}(\varDelta t_{m}).
\end{equation}
By partially tracing over the degree of the electron spin, the final
state of the nuclear spin $\rho^{(T)}(\tau)=\textrm{Tr}_{D}\left[\rho(\tau)\right]$
reads
\begin{equation}
\rho^{(T)}(\tau)=\left(\begin{array}{cc}
\alpha & e^{-i\gamma_{n}B_{z}\tau}\cos(\varOmega\varDelta t_{m})\beta\\
e^{i\gamma_{n}B_{z}\tau}\cos(\varOmega\varDelta t_{m})\beta^{*} & 1-\alpha
\end{array}\right).
\end{equation}

On the other hand, the initial state of the nuclear spin can be decomposed into its eigenbasis as \cite{Layden15}
\begin{eqnarray}
\rho^{(T)}(0)=C_{0}I+C_{1}\sigma_{+}+C_{2}\sigma_{-}+C_{3}\sigma_{z},
\end{eqnarray}
where $C_{0}=1/2$, $C_{1}=\beta$, $C_{2}=\beta^{*}$, $C_{3}=\left(2\alpha-1\right)/2$,
$I$ is the identity operator, $\sigma_{+}=\left|\uparrow\rangle\langle\downarrow\right|$
and $\sigma_{-}=\left|\downarrow\rangle\langle\uparrow\right|$ are the raising and
lowering operators respectively. After the first cycle, the nuclear spin is in the state
\begin{equation}
\rho^{(T)}(\tau)=C_{0}\lambda_{0}I+C_{1}\lambda_{1}\sigma_{+}+C_{2}\lambda_{2}\sigma_{-}+C_{3}\lambda_{3}\sigma_{z}\label{eq:rho-tau-1},
\end{equation}
where the eigenvalues are
\begin{align}
\lambda_{0}&= \lambda_{3}=1\label{eq:Lambda03},\\
\lambda_{1}&= \lambda_{2}^{*}=e^{-i\gamma_{n}B_{z}\tau}\cos(\varOmega\varDelta t_{m}) \label{eq:Lambda}.
\end{align}
After $N$ cycles, the nuclear-spin qubit evolves into the state
\begin{equation}
\rho^{(T)}(N\tau)=C_{0}\lambda_{0}^{N}I+C_{1}\lambda_{1}^{N}\sigma_{+}+C_{2}\lambda_{2}^{N}\sigma_{-}+C_{3}\lambda_{3}^{N}\sigma_{z}.
\label{eq:Ncircle}
\end{equation}
Here, $\lambda_{0}=\lambda_{3}=1$ are related to fixed points \cite{Brouwer11} independent
of all parameters. However, $\lambda_{1}$ and $\lambda_{2}$ are modulated
by the parameters ($\varOmega$, $B_{z}$, $\varDelta t_{m}$, $\varDelta t_{f}$).
By adjusting these parameters, the quantum Zeno and Zeno-like effects can be observed.

Hereafter, by analyzing the dependence of the eigenvalues on the parameters,
we demonstrate the existence of quantum Zeno and Zeno-like effects.

\subsubsection*{Quantum Zeno-like effect}

In equation~(\ref{eq:rho-tau-1}), the eigenvalues $\lambda_{0}=\lambda_{3}=1$
mean that $\rho^{(T)}(0)=C_{0}I+C_{3}\sigma_{z}$ are the
fixed points independent of the combination of parameters ($\varOmega$, $B_{z}$,
$\varDelta t_{m}$, $\varDelta t_{f}$) after repeated measurements.
To be specific, if the initial state of the nuclear spin is of diagonal form, the state will not change and thus is preserved.
This is a QZLE on the nuclear spin, similar to that in Ref.~\citen{Layden15}.

On the other hand, the $\sigma_{+}$ and $\sigma_{-}$ components
of the initial nuclear-spin state are exponentially suppressed, when $\left|\lambda_{1}|=|\lambda_{2}\right|<1$
with appropriate parameters ($\varOmega$, $B_{z}$, $\varDelta t_{m}$,
$\varDelta t_{f}$). Therefore, the following process
$\rho^{(T)}(0)=C_{0}I+C_{1}\sigma_{+}+C_{2}\sigma_{-}+C_{3}\sigma_{z}\rightarrow\rho^{(T)}(N\tau)=C_{0}I+C_{3}\sigma_{z}$
is achieved by sufficiently-many measurements.

Furthermore, the existence of $\lambda_{3}=1$ preserves the polarization of
the nuclear spin. If $\alpha$ is 1 or 0,
$\rho^{(T)}(N\tau)=\rho^{(T)}(0)=\left|\uparrow\rangle\langle\uparrow\right|$ or
$\rho^{(T)}(N\tau)=\rho^{(T)}(0)=\left|\downarrow\rangle\langle\downarrow\right|$ can be obtained.
Thus, the polarization of the $^{13}\mathrm{C}$ nuclear spin
near the NV center is frozen. It may be a potential way to
preserve the polarization of $^{13}\mathrm{C}$
nuclear spin against its hyperfine interaction with electron spin.

\subsubsection*{Quantum Zeno effect}

For the eigenvalues $\lambda_{1}$ and $\lambda_{2}$, we
consider both the ideal situation when $\varDelta t_{f}$, $\varDelta t_{m}$$\rightarrow0$
and the realistic situation of finite $\varDelta t_{f}$ and $\varDelta t_{m}$.

In the ideal situation, equation~(\ref{eq:Lambda}) is simplified as
\begin{align}
\lambda_{1} = \lambda_{2}^{*}\approx e^{-i\gamma_{n}B_{z}\tau}\left(1-\frac{\varOmega^2}{2}\varDelta t_{m}^{2}\right).
\end{align}
From equation~(\ref{eq:Ncircle}), we learn that the eigenvectors $\sigma_{+}$ and $\sigma_{-}$ contribute to the off-diagonal part of $\rho^{(T)}(\tau)$. When $\varDelta t_{f}$ and
$\varDelta t_{m}$ are small enough, the eigenvalues $\left|\lambda_{1}\right|$
and $\left|\lambda_{2}\right|$ can be very close to unity.
Furthermore, $\gamma_{n}B_{z}N\tau=2n\pi$ is assumed, where $N\tau=T$ is the fixed total evolution time and \emph{n} is an integer. Because the eigenvalues $\left|\lambda_{1}\right|$
and $\left|\lambda_{2}\right|$ approach unity quadratically
in the high-measurement-frequency limit, an arbitrary nuclear-spin state
is exactly preserved by infinitely-frequent measurements.
Here, the conventional QZE is recovered.

\begin{figure}[ht]
\centering
\includegraphics[scale=0.5]{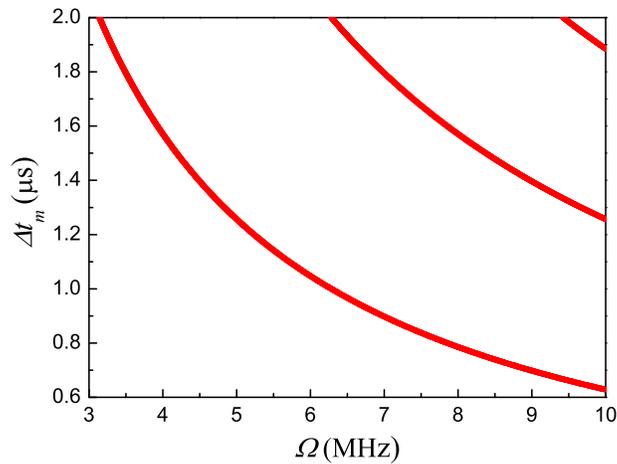}
\caption{Measurement conditions for the QZLE.
The QZLE will occur at the specified locations in the
parameter space ($\varOmega$, $\varDelta t_{m}$). The points are determined
by equation~(\ref{eq:condition}) with $n_{1}=$ 1, 2, 3\label{fig:MeasureCon}.}
\end{figure}

On the other hand, consider the realistic condition in an NV center, e.g. non-vanishing $\varDelta t_{m}$ due to a finite pulse width.
When the Rabi frequency and the pulse width are chosen to meet the following requirement
\begin{equation}
\varOmega\varDelta t_{m}=2n_{1}\pi \label{eq:condition}
\end{equation}
with $n_{1}$ being positive integer, $\lambda_{1}=\lambda_{2}=1$ can be obtained.
In this case, arbitrary initial state is the Zeno-like fixed point
which depend on the appropriate choice of parameters
($\varOmega$, $B_{z}$, $\varDelta t_{m}$, $\varDelta t_{f}$).
In other words, under certain measurement conditions,
the QZLE is observed. Figure~\ref{fig:MeasureCon} shows the locations where QZLE
will occur in the parameter space of ($\varOmega$, $\varDelta t_{m}$).
In the case of $\left|\lambda_{1}\right|=\left|\lambda_{2}\right|<1$, after repeated measurements the elements of $\sigma_{+}$ and $\sigma_{-}$ in $\rho^{(T)}(T)$ may disappear due to accumulated loss.
However, by tuning parameters to the vicinity of the points ($\varOmega$, $\varDelta t_{m}$) given in equation~(\ref{eq:condition}), equation~(\ref{eq:Lambda}) can be expanded to the second order of $\varDelta t_{m}$ as
\begin{equation}
\lambda_{1}=\lambda_{2}^{*}\approx1-\frac{1}{2}(\varOmega\varDelta t_{m}-2n_{1}\pi)^{2}.
\end{equation}
Here, the conventional QZE happens in the neighbourhood of the QZLE points.
In other words, the QZE occurs for a series of parameter combinations
corresponding to finite-frequency measurements with finite coupling strengths.
Since the eigenvalues $\lambda_{1}$ and $\lambda_{2}$ approach unity
quadratically under the repeated measurements, any nuclear-spin state
is exactly preserved by finitely-frequent measurements, even though
it is affected by the free evolution.

\section*{Discussion}

The conventional QZE and QZLE are demonstrated in a $\mathrm{^{13}C}$ nuclear spin
around an NV center by controlling the electron spin.
Both of the QZE and QZLE can be observed by modulating
the Rabi frequency, and the magnetic field, and the free-evolution time, and the pulse width.
Our numerical calculation properly shows that for suitable parameters
there exist both the QZE and QZLE in an NV-center system
under the experimental condition. Consequently, the conventional
QZE and QZLE are obtained with finite-frequency imperfect measurements.

In order to put our experimental proposal into practice,
the secular approximation should be valid, i.e. the applied magnetic field
strength should not be too strong \cite{Dreau13}. As a consequence,
the magnetic field strength $B_{z}$  could be less than 200 G.
Additionally, due to the resonance condition, the driving frequency $\omega$  equals to
the level spacing between $\left|0,\uparrow\right\rangle$ and $\left|-1,\uparrow\right\rangle$, i.e., $D-\gamma_{e}B_{z}-A_{zz}/2$. At the same time, the level spacing
$\omega_{1}$ between $\left|0,\downarrow\right\rangle$ and $\left|-1,\downarrow\right\rangle$
is $D-\gamma_{e}B_{z}+A_{zz}/2$. To selectively only induce the transition between
$\left|0,\uparrow\right\rangle$ and $\left|-1,\uparrow\right\rangle$, the large-detuning condition
$\varDelta \omega=\omega_{1}-\omega =A_{zz} \gg \varOmega$ should be fulfilled.
Since the hyperfine coupling between the electron spin and a $\mathrm{^{13}C}$ nuclear spin
in the first coordination shell is known to be 130 MHz \cite{Jelezko04,Smeltzer11},
the Rabi frequency $\varOmega$ can be no more than 10 MHz.
Furthermore, the initialization of the NV center will take approximately $\sim\!140$~ns \cite{Dutt07}, and $\varDelta t_{m}$ and $\varDelta t_{f}$ are on a time scale about $2~\mu \textrm{s}$. Thus, a single cycle process will take about $5~\mu \textrm{s}$. The intrinsic dephasing time of the $\mathrm{^{13}C}$
nuclear spin $T_{2n}$ was observed as  around one second \cite{Maurer12}.
To ignore the decoherence effect induced by the environment, we restrict the total
experiment time as $T\ll T_{2n}$, i.e., the total experiment time \emph{T} is chosen as %$50~\mu \textrm{s}$.
$100$ ms. In this case, we can demonstrate the quantum Zeno and Zeno-like effects for roughly $2 \times 10^4$ cycles.

On the other hand, due to the presence of the nitrogen nucleus, $^{14}\mathrm{N}$ (\emph{I}=1) or $^{15}\mathrm{N}$ (\emph{I}=1/2), and the $^{13}\mathrm{C}$ nuclear spin bath,
the dephasing time of the electron spin is $58~\mu \textrm{s}$.\cite{Kennedy03}
The duration $\tau=\varDelta t_{f}+\varDelta t_{m}$ of a cycle is smaller
than the dephasing time of the electron spin by one order. Because the hyperfine coupling
between the electron spin and the nitrogen nucleus  $A_{N}<4$ MHz \cite{Felton09}
is much smaller than $A_{zz}$, the dephasing effect induced by the nitrogen nucleus can be neglected.
The dipole-dipole interactions between the electron spin and the other $^{13}\mathrm{C}$
nuclear spins are too weak. Therefore, the dephasing effect induced by all nuclear spins can also be neglected.

Last but not the least, the nuclear spin bath induces the dephasing of the $\mathrm{^{13}C}$ nuclear spin. Because the total experiment time is sufficiently short and the magnetic
dipole-dipole interactions between the $\mathrm{^{13}C}$ nuclear spin and
the other nuclear spins are weak enough \cite{Bermudez11}, the dephasing effect induced by
the spin bath can be neglected. Meanwhile, after every measurement, we decouple the
electron and $\mathrm{^{13}C}$ nuclear spins and initialize the electron spin in
its ground state $|0\rangle$ \cite{Dutt07,Jiang08} without
perturbing the $\mathrm{^{13}C}$ nuclear spin. In this process,
the nuclear spin is supposed to be completely isolated from the environment \cite{Dutt07,Jiang08}.
Therefore, the nuclear spin hardly evolves during this process.

In conclusion, as shown in Fig.~\ref{fig:MeasureCon}, our numerical
calculation properly indicates that under practical conditions we can demonstrate
the conventional QZE and QZLE in $\mathrm{^{13}C}$ nuclear spin
around the NV center with finite-frequency imperfect measurements.

\section*{Methods}

\textbf{The rotating frame.} Since the original Hamiltonian in the measurement process $H_M$ is time-dependent, the whole system is transformed to a rotating frame defined by the transformation
$\bigr|\varPsi(t)^{R}\bigl\rangle=U^{\dagger}(t)\bigr|\varPsi(t)\bigl\rangle$, where $U(t)=\exp(-i\xi t)$, $\bigr|\varPsi(t)\bigl\rangle$ and $\bigr|\varPsi(t)^{R}\bigl\rangle$ are respectively the wave functions in the static and rotating frames. Now, we derive the relationship between the Hamiltonian in the static frame $H_{M}$ and the Hamiltonian in the rotating frame $H_{M}^{R}$. Because the time evolution of $\bigr|\varPsi(t)^{R}\bigl\rangle$ still fulfills the Schrodinger equation in the rotating frame, i.e.
\begin{align}
i\dot{\bigr|\varPsi(t)^{R}\bigl\rangle}	&=	i\frac{d}{dt}U^{\dagger}\bigr|\varPsi(t)\bigl\rangle       \nonumber \\
	&=	i\dot{U^{\dagger}}\bigr|\varPsi(t)\bigl\rangle+U^{\dagger}i\dot{\bigr|\varPsi(t)\bigl\rangle}      \nonumber  \\
	&=	i\dot{U^{\dagger}}UU^{\dagger}\bigr|\varPsi(t)\bigl\rangle+U^{\dagger}H_{M}UU^{\dagger}\bigr|\varPsi(t)\bigl\rangle       \nonumber \\
	&=	i\dot{U^{\dagger}}U\bigr|\varPsi(t)^{R}\bigl\rangle+U^{\dagger}H_{M}U\bigr|\varPsi(t)^{R}\bigl\rangle      \nonumber  \\
	&=	(i\dot{U^{\dagger}}U+U^{\dagger}H_{M}U)\bigr|\varPsi(t)^{R}\bigl\rangle      \nonumber  \\
	&=	(U^{\dagger}H_{M}U-\xi)\bigr|\varPsi(t)^{R}\bigl\rangle,
\end{align}
the effective Hamiltonian in the rotating frame reads
\begin{equation}
H_{M}^{R}=U^{\dagger}H_{M}U-\xi. \label{eq:rotating-frame}
\end{equation}
Correspondingly, the relationship between the density matrix in the static frame $\rho(t)$
and the density matrix in the rotating frame $\rho^{R}(t)$ is
\begin{align}
\rho(t)	&=	\bigr|\varPsi(t)\bigl\rangle\bigr\langle\varPsi(t)\bigr|      \nonumber \\
	&=	U(t)\bigr|\varPsi(t)^{R}\bigl\rangle\bigr\langle\varPsi(t)^{R}\bigr|U^{\dagger}(t)   \nonumber \\
	&=	U(t)\rho^{R}(t)U^{\dagger}(t). \label{eq:rotating-frame-state}
\end{align}

%\textbf{Single cycle.}

\section*{Acknowledgements}

FGD was supported by the National Natural Science Foundation of China under Grant No.~11474026 and
the Fundamental Research Funds for the Central Universities under Grant No.~2015KJJCA01. QA was supported
by the National Natural Science Foundation of China under Grant No.~11505007,
the Youth Scholars Program of Beijing Normal University under Grant No.~2014NT28,
and the Open Research Fund Program of the State Key Laboratory of Low-Dimensional
Quantum Physics, Tsinghua University Grant No.~KF201502. MZ was supported by the National Natural Science Foundation of China under Grant No.~11475021. ZQY is funded by the
National Key Basic Research Program of China under Grant
No.~2011CBA00300 and the National Natural Science Foundation
of China under Grant Nos.~11105136 and~61435007.

\section*{Author contributions statement}

Q.A., Z.Q.Y., F.G.D., M.Z., J.Q. and Y.Y.W. wrote the main manuscript text,
J.Q. and Y.Y.W. did the calculations. Q.A., F.G.D, M.Z. and Z.Q.Y. designed the project.
Q.A. supervised the whole project. All authors reviewed the manuscript.

\section*{Additional information}

Competing financial interests: The authors declare no competing financial interests.

\end{document}